# Fully Automated Mitral Inflow Doppler Analysis Using Deep Learning

Mohamed Y. Elwazir
Department of Cardiovascular Medicine, *Mayo Clinic*
Rochester, MN, USA
Elwazir.Mohamed@mayo.edu
Cardiology Department
Suez Canal University
Ismailia, Egypt

Zeynettin Akkus
Department of Cardiovascular Medicine, *Mayo Clinic*
Rochester, MN, USA
akkus.zeynettin@mayo.edu

Zi Ye
Department of Cardiovascular Medicine, *Mayo Clinic*
Rochester, MN, USA
ye.zi@mayo.edu

Didem Oguz
Department of Cardiovascular Medicine, *Mayo Clinic*
Rochester, MN, USA
Oguz.Didem@mayo.edu

Jae K. Oh
Department of Cardiovascular Medicine, *Mayo Clinic*
Rochester, MN, USA
Oh.Jae@mayo.edu

*Abstract*—Echocardiography (echo) is an indispensable tool in a cardiologist's diagnostic armamentarium. To date, almost all echocardiographic parameters require time-consuming manual labeling and measurements by an experienced echocardiographer and exhibit significant variability, owing to the noisy and artifact-laden nature of echo images. For example, mitral inflow (MI) Doppler is used to assess left ventricular (LV) diastolic function, which is of paramount clinical importance to distinguish between different cardiac diseases. In the current work we present a fully automated workflow which leverages deep learning to a) label MI Doppler images acquired in an echo study, b) detect the envelope of MI Doppler signal, c) extract early and late filing (E- and A-wave) flow velocities and E-wave deceleration time from the envelope. We trained a variety of convolutional neural networks (CNN) models on 5544 images of 140 patients for predicting 24 image classes including MI Doppler images and obtained overall accuracy of 0.97 on 1737 images of 40 patients. Automated E and A wave velocity showed excellent correlation (Pearson R 0.99 and 0.98 respectively) and Bland-Altman agreement (mean difference 0.06 and 0.05 m/s respectively and SD 0.03 for both) with the operator measurements. Deceleration time also showed good but lower correlation (Pearson R 0.82) and Bland-Altman agreement (mean difference: 34.1ms, SD: 30.9ms). These results demonstrate feasibility of Doppler echocardiography measurement automation and the promise of a fully automated echocardiography measurement package.

## I. Introduction

Echocardiography is among the most versatile and ubiquitous diagnostic tools in cardiology. Aside from two- and three-dimensional imaging of cardiac structure and function, it is unrivaled as a rapid and noninvasive means to assess cardiac hemodynamics. By making use of Doppler ultrasound, the speed and direction of blood flow at a given point or line can be determined and visualized, allowing quantitative and reproducible assessment of various aspects of myocardial and valvular function. Mitral inflow pulsed-wave Doppler is one of the most well-established methods for evaluating left ventricular diastolic function, with the most common parameters utilized being E and A wave amplitude, E/A ratio, and E wave deceleration time [1]. Such Doppler measurements are normally obtained through the time-consuming approach of manual tracing/marking of the Doppler envelope by a qualified echocardiographer.

### A. Challenges

Aside from being a time-consuming procedure that requires a skilled operator, Doppler measurements exhibit significant inter- and intra-operator variability [2]–[4]. To overcome this issue, measurements from multiple beats and more than one acquisition are usually obtained and averaged, adding to the time and resources required.

Furthermore, Doppler signals usually suffer from artifacts such as aliasing in pulsed-wave tracings and contamination (spillover of flow from regions/valves other than the one being interrogated). In such cases it is up to the sonographer to apply his knowledge and experience to discriminate and exclude the artifactual signal from processing. Automation approaches which rely on image thresholding and conventional signal processing (see section II.B) are vulnerable to such artifacts, especially contamination.

### B. Current solutions and their drawbacks

The most commonly used approach for Doppler trace extraction relies on digital image processing in the form of a noise filtering process, followed by some form of thresholding and an edge detection algorithm [5]. Signal processing is then used to obtain the measurements. This method is, however, liable to contamination artifacts and is heavily dependent on image quality. Faint parts of the Doppler trace might get falsely filtered out by the thresholding process, especially in the presence of a noisy background.

Active contour modeling (snakes) is another approach which entails curve detection using an optimization process involving elastodynamic models of curve contrast and smoothness [6]. It has been applied in various fields including Doppler echocardiography but is also liable to the same quality issues as digital image processing.

A probabilistic, hierarchical and discriminant framework was proposed by Zhou *et al*. [7] for detection of deformable anatomic structures from medical images including Doppler traces. This machine-learning based algorithm relies on detection of key fiducial points in the trace and recursively merging them to create a simplified skeleton of the trace used to derive the measurements. This simplification, however, leads to loss of information and a lower overall accuracy.

These approaches are all semi-automated, in the sense that they require manual selection of the images to be processed for each workflow.

*C. Aims*

In the current study we present an example of a fully automated pipeline for mitral inflow Doppler analysis using a combination of deep learning classification/segmentation and signal processing. The proposed workflow is able to learn to handle image quality issues or artifacts and can, with minor modifications, be adapted to work with any echocardiographic Doppler tracing.

## II. METHODS

*A. Patient Population*

The study was approved by the Mayo Clinic institutional review board. Pulsed wave mitral inflow Doppler images were collected from patients referred for echocardiographic examination in the Echocardiography Department at Mayo clinic between November 2019 and February 2020. All patients consented to research authorization. Only patients with normal echocardiograms were included, defined as normal ventricular size and function and no significant (more than mild) valvular disease.

*B. Data Collection*

We downloaded 200 transthoracic exams (TTE) from the imaging archive of Mayo echocardiography lab. The TTE exams were performed using GE Vivid E95 (GE Healthcare, US). In order to obtain mitral inflow velocity Doppler signal, an apical four chamber view is obtained first and the pulsed-wave Doppler cursor is positioned at the tips of the mitral valve leaflets as recommended in guidelines [8]. The operators performing the exam adjust the images at their discretion and obtain the standard measurements from Doppler velocity signal: E wave and A wave peak amplitudes and deceleration time defined as the time from the E wave peak to where the wave slope meets the baseline. Still RGB images with and without the operator measurements are saved as DICOM files. The Doppler images we used in this study have the size of 1016×758 pixels.

*C. Analysis*

Our fully automated pipeline for mitral inflow Doppler images includes several steps: image preprocessing, image classification, Doppler signal envelope detection, ECG extraction and QRS peak detection. The flowchart of the automated pipeline is shown in Figure 1.

*1) Image Preprocessing*

The gray scale values of images that range from 0 to 255 were divided by the maximum gray scale value of 255 to scale all values between 0 to 1. All images are resized to 299-by-299 pixels using bi-linear interpolation [9] for the image classification task. For the envelope detection task, we zero padded each image to a larger standard bounding box in size of 1024x1024.

*2) Image classification*

We used five well-known CNN architectures, which are inception [10], [11], resnet50 [12], densenet121 [13], inception_resnet [14], and VGG16 [15], to classify all still images in an echo study into 24 classes. Only still images were used for the classification task. Cine clips containing multiple image frames were excluded. We defined 24 classes of still images (figure 2, see appendix 1 for full names), We used Adam optimizer with a learning rate of 1e-4 during the training. The models were trained until a steady state was reached.

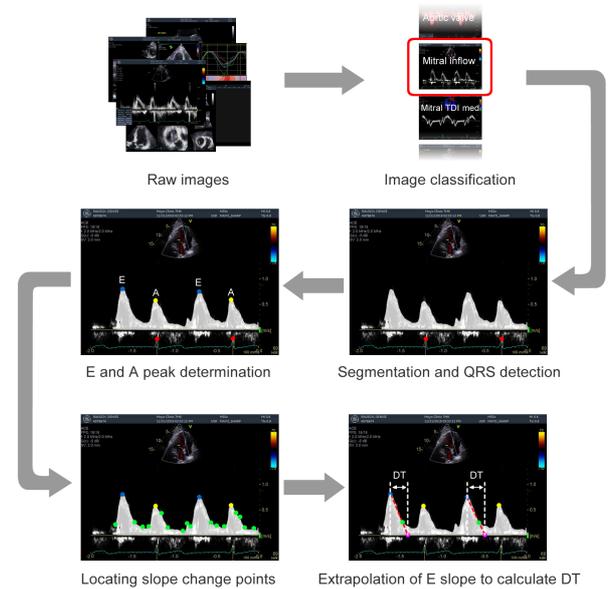

Figure 1: The flowchart of fully automated mitral inflow Doppler analysis.

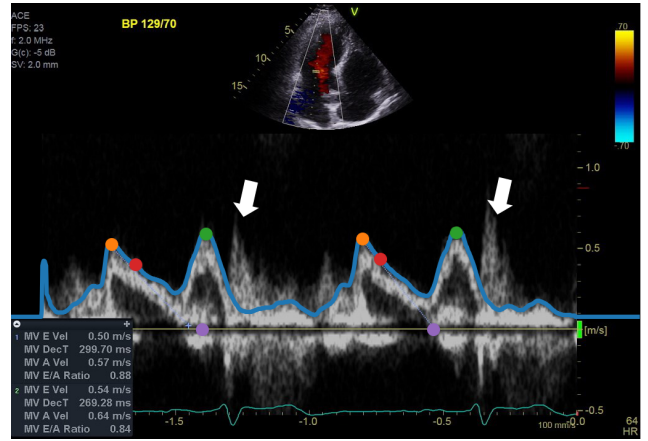

Figure 2. Example automatically processed Doppler mitral inflow trace. The blue line indicates the automated segmentation upper border. It can be noted how the learning-based approach excludes bright artifactual peaks (arrows) which thresholding would have falsely included. Orange points indicate detected E waves while green points indicate detected A waves. Red points indicate points of change in E wave slope, and magenta points are the extrapolation of the slope to the baseline. Deceleration time is defined as the duration (horizontal axis distance in milliseconds) between the E wave peak (orange point) and the end of the slope (magenta points).

*3) Doppler signal envelope detection*

We trained an encoder-decoder type convolutional neural networks (CNN) model which adapts the U-Net architecture [16] The model was trained to segment the image into the Doppler signal from baseline to the envelope (while avoiding aliasing and other artifacts) with everything else as background (Figure 2). The model was initialized with Glorat uniform distribution [17] and used a softmax classifier, which outputs normalized probabilities of two classes. The model is capable of receiving any input size and pads with zeros to standardize the input image size to 1024-by-1024 pixels without changing the aspect ratio of the input image. The output of the model is two classes with the same size as the input image (2x1024x1024). Since the number of samples (pixels) of the background class is much higher than that of the mitral inflow Doppler signal, we used weighted categorical crossentropy loss [18]–[20] during training to balance the impact of the two classes. We also used Adam optimizer with a learning rate of 1e-4 during the training. The model was trained on 100 mitral inflow Doppler images from 30 patients. The reference envelope of the Doppler signal was manually segmented by a cardiologist (MYE) to train the model. In the final workflow, the image classification and envelope detection models run in sequence to find mitral inflow Doppler images and trace the envelope.

*4) ECG extraction and QRS detection*

The raw image was processed to extract the embedded electrocardiogram (ECG) tracing based on pixel color, and signal processing was used to obtain the QRS complex peaks which correspond to ventricular electric activation.

*5) Measuring early and late filing flow velocities and deceleration Time*

Signal processing was used to find the peaks of early and late filing flow velocities (E and A waves) of the segmented envelope of Doppler flow signal, while setting minimum requirements for peak width and prominence to avoid capturing noise peaks. The ECG QRS data from the previous step was used to discern the E and A waves (the A wave immediately precedes the QRS complex). Doppler velocity scale information was obtained from the DICOM file when available or using Optical character recognition (OCR) on the image otherwise. This was used to calculate E and A velocities from their pixel amplitudes.

Deceleration time was calculated by obtaining the second derivative of the smoothed doppler trace border to obtain points of change in slope, then extending a line from the E peak to the next point of slope change, extrapolating the line to the baseline, and measuring the horizontal distance in pixels which was then converted to time in milliseconds using the embedded DICOM time scale data. This process was repeated for all available envelopes. Figure 2 shows an example.

D. *Experiments and evaluations*

In the image classification task, we divided our dataset (i.e. 200 patients) based on patient level into 70% (5544 images of 140 patients) for the training, 10% (811 images of 20 patients) for the validation, and 20% (1737 images of 40 patients) for the testing purposes. We assessed the performance of the five well-known CNN models (i.e. inception, resnet50, densenet, inception_resnet, and VGG16) to choose the best performing model for our classification task.

Three experts manually annotated and measured E and A velocity as well as deceleration time. Expert 1 annotated 96 beats from 48 patients for comparison with the automated measurements, while experts 2 and 3 annotated only 20 images to assess interoperator variability.

The automated results were compared to manually obtained results from expert 1. Pearson correlation and coefficient of determination (linear regression R squared) were used to assess agreement between the automated measurements and the human operators. Bland-Altman analysis was done to calculate mean difference (bias) and standard deviation (SD) with a confidence interval of ±2SD. Data analysis was performed in R version 3.5.1 using the BlandAltmanLeh package.

III. RESULTS

As shown in Table 1 and 2, the inception model with residual connections gave the best performance on the test dataset to classify images into 24 classes. Mitral inflow images both in validation and test sets were identified with high accuracy and 100% sensitivity as shown in Figure 3 that includes the confusion matrix of all classes.

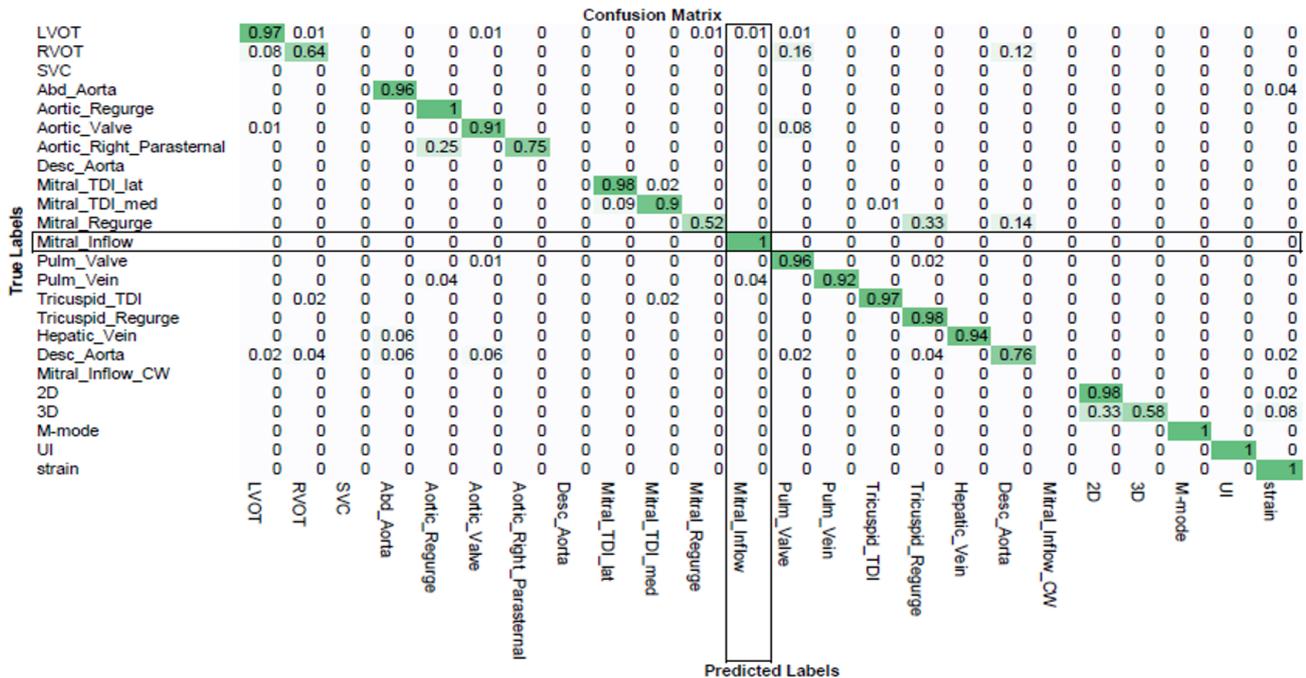

Figure 3: Confusion matrix for the image classification performed on the test dataset. The values are normalized along the predicted label axis.

Table 1: The performance of five CNN models for the image classification on the validation dataset.

| Validation dataset | Accuracy | Precision | Recall | f1-score |
|---|---|---|---|---|
| Inception | 0.96 | 0.89 | 0.87 | 0.88 |
| Resnet50 | 0.96 | 0.93 | 0.90 | 0.90 |
| Inception_Resnet | 0.97 | 0.96 | 0.90 | 0.92 |
| DenseNet121 | 0.97 | 0.96 | 0.92 | 0.94 |
| VGG16 | 0.95 | 0.94 | 0.90 | 0.91 |

Table 2: The performance of five CNN models for the image classification on the test dataset.

| Test Dataset | Accuracy | Precision | Recall | f1-score |
|---|---|---|---|---|
| Inception | 0.94 | 0.86 | 0.84 | 0.85 |
| Resnet50 | 0.93 | 0.83 | 0.78 | 0.79 |
| Inception_Resnet | 0.95 | 0.93 | 0.89 | 0.90 |
| DenseNet121 | 0.95 | 0.89 | 0.84 | 0.86 |
| VGG16 | 0.91 | 0.87 | 0.81 | 0.83 |

Table 3 shows performance metrics (bias, Pearson R and linear regression $R^2$) for the automated system versus human operators, while Figure 4 shows scatter and Bland-Altman plots. Mean difference (bias) for the automated measurements versus the operator measurements was 0.06 m/s for E wave, 0.05 and A waves and 37.2 ms for DT. Pearson R and $R^2$ for E and A waves were in the high nineties indicating excellent agreement. Pearson R for DT was 0.82 which despite still indicating a strong correlation, was lower than that for E and A waves since any small change in slope leads to a significant difference in millisecond DT duration.

Figure 5 shows the results of interobserver variability analysis. For 20 patients, the average for each patient's beats was calculated and plotted for the three operators as well as the automated measurements.

## I. Discussion

In this study, we demonstrate the superiority and versatility of a machine learning based approach to Doppler trace analysis. A fully automated pipeline was presented which can identify mitral inflow Doppler images, perform segmentation of the Doppler flow signal while avoiding artifacts, and then use signal processing to extract relevant measurements. Correlation between the automated measurements and the human operator measurements was shown to be good using Pearson correlation, linear regression, and Bland-Altman analysis. This workflow shows that most measurements can be similarly automated, with the concept of a fully automated echocardiographic quantification package being within increasingly close reach.

A small positive bias is observed in the automated measurements (Table 3, Figures 4 and 5). This is attributable to the fact that a sonographer will often ignore the fainter upper border of the trace and use the line of densest signal directly below to mark the peak. The segmentation network, on the other hand, was trained to include the trace in its entirety. This bias, however, is of the systematic nature, as evidenced by its elimination when dividing the E by A amplitude in the E/A ratio (Figure 5). This could be further improved by adjusting manual segmentations that were used for training the model.

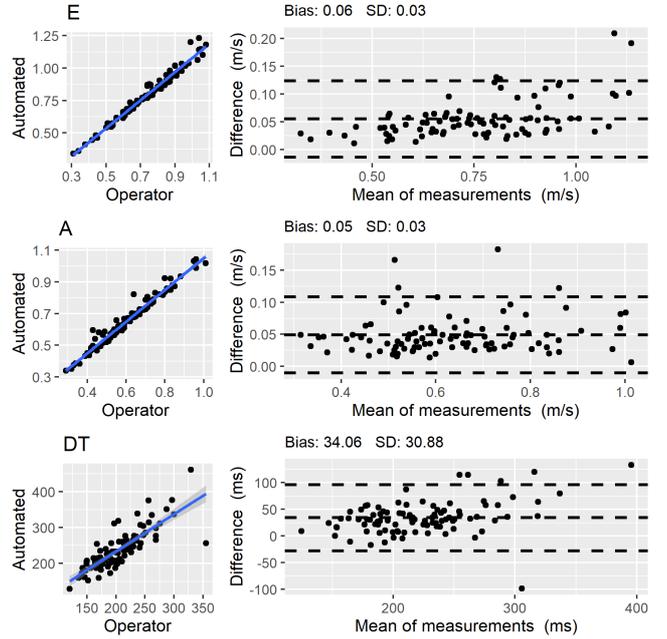

Figure 4. Scatter plots (left) and Bland-Altman plots (right) for automated vs. operator measurements of E wave amplitude (top), A wave amplitude (middle) and DT (bottom). SD: standard deviation; DT: deceleration time.

TABLE 3. PERFORMANCE METRICS OF THE AUTOMATED SYSTEM

| Parameter | Bias | Pearson R | $R^2$ |
|---|---|---|---|
| E wave (m/s) | 0.06 | 0.986 | 0.972 |
| A wave (m/s) | 0.05 | 0.982 | 0.965 |
| Deceleration time (ms) | 34.06 | 0.82 | 0.672 |

The main limitation of this study is that the classification and segmentation networks were trained exclusively on normal echocardiograms from a single vendor (GE). Further training and validation on multi-vendor images obtained from a range of normal and abnormal patients is required as a next step.

Another limitation is that the network, unlike human echocardiographers, is unable to detect and discard low quality beats from the assessment. Since these beats will likely have significantly different measurements, a potential solution is to exclude beats with outlying E, A or DT values.

Our workflow uses ECG to discriminate the E and A waves as opposed to other studies that use autocorrelation [21]. This has the disadvantage of being dependent on the presence and quality of the accompanying ECG trace. However, in the case of the dual-peaked mitral inflow signal, the autocorrelation method is likely to fail with more rapid heart rates where systolic and diastolic durations become equal and E and A waves become evenly spaced.

In summary, this fully automated workflow allows faster and consistent analysis of larger amounts of imaging data without human intervention. Operators will frequently need to perform several measurements and average them in order to reduce discrepancies introduced by respiration, small changes in probe position, and beat-to-beat variations in cardiac cycle length. Automated analysis allows rapid analysis of any number of beats acquired with significant reduction in examination time, allowing more examinations to be performed per machine and more importantly reducing the variability in the measurements of operators.

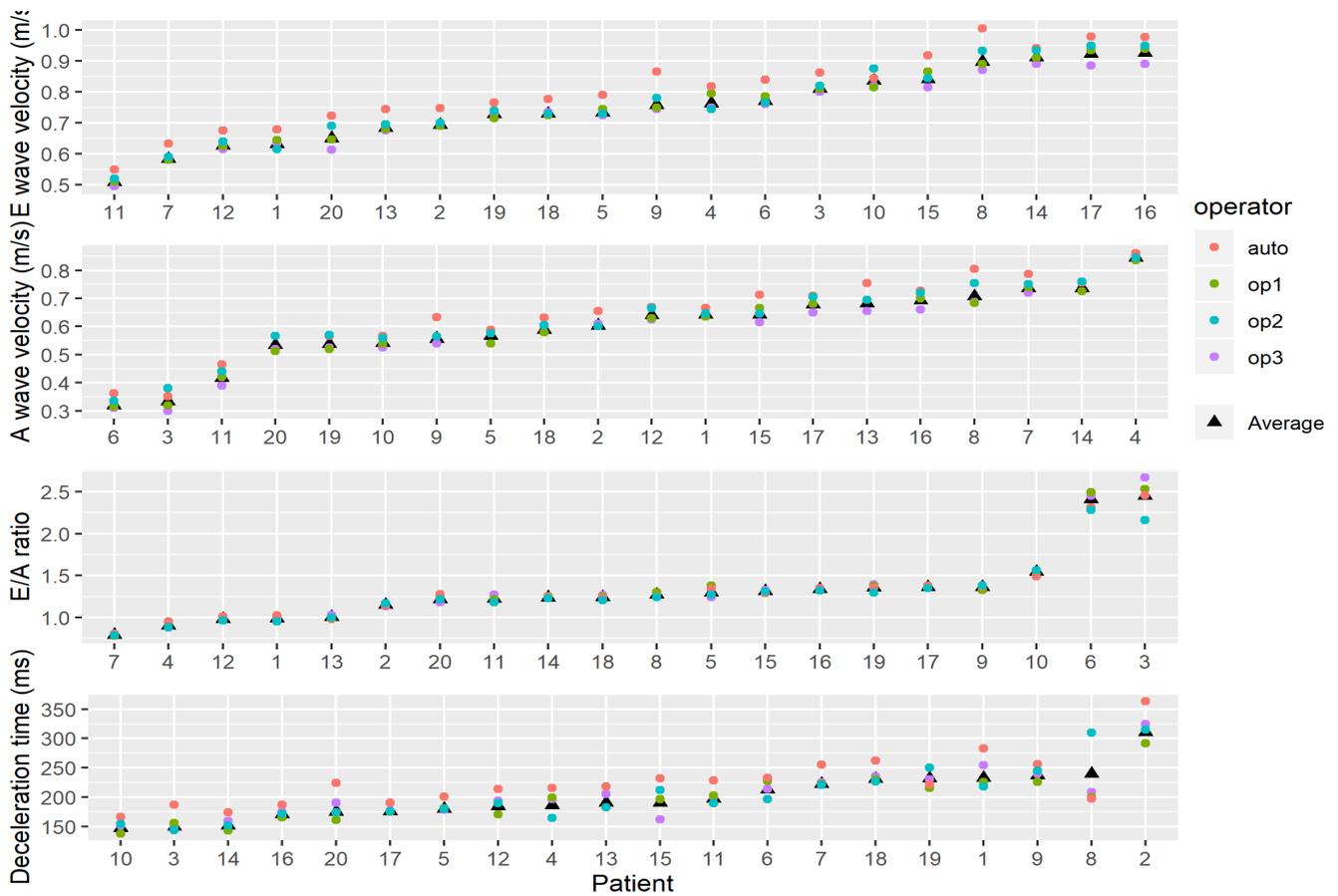

Figure 5. Scatter plot of patient-wise all-beat average for each of three operators and automated measurements. The average of the three operators is represented by black triangles.

APPENDIX A

Table A1 shows the full list of classes for the image classification network:

| Label | Full name |
| --- | --- |
| LVOT | Left ventricular outflow tract Doppler |
| RVOT | Left ventricular outflow tract Doppler |
| SVC | Superior vena cava Doppler |
| mitral_inflow_CW | mitral inflow continuous wave Doppler |
| abd_aorta | Abdominal aorta Doppler |
| aortic_regurge | Aortic regurgitation Doppler |
| aortic_right_parasternal | Aortic flow Doppler right parasternal position |
| aortic_valve | Aortic outflow Doppler |
| desc_aorta | Descending aorta Doppler |
| hepatic vein | Hepatic vein Doppler |
| mitral_TDI_lat | Mitral lateral annulus tissue Doppler |
| mitral_TDI_med | Mitral medial annulus tissue Doppler |
| mitral_inflow | Mitral inflow pulsed wave Doppler |
| mitral_regurge | Mitral regurgitation |
| pulm_valve | Pulmonary valve Doppler |
| pulm_vein | Pulmonary vein Doppler |
| tricuspid_regurge | Tricuspid regurgitation |
| tricuspid_TDI | Mitral annulus tissue Doppler |
| hepatic_vein | Hepatic vein Doppler |
| 2D | 2-dimensional image |
| 3D | 3-dimensional image |
| UI | User interface screen capture |
| strain | Strain map |